\documentclass[twocolumn,superscriptaddress,prb,aps]{revtex4}
\usepackage{amsmath}
\usepackage{amssymb}
\usepackage{graphicx}
\usepackage{dcolumn}

\begin{document}

\title{Resonant optical control of the electrically-induced spin polarization \\ by periodic excitation}
\author{F. G. G. Hernandez}
\email[Corresponding author.\\ Electronic address:]{felixggh@if.usp.br} \affiliation{Instituto de F\'isica, Universidade de S\~{a}o Paulo, Caixa Postal 66318 -
CEP 05315-970, S\~{a}o Paulo, SP , Brazil}
\author{G. M. Gusev}
\affiliation{Instituto de F\'isica, Universidade de S\~{a}o Paulo, Caixa Postal 66318 -
CEP 05315-970, S\~{a}o Paulo, SP , Brazil}
\author{A. K. Bakarov} \affiliation{Institute of Semiconductor Physics, Novosibirsk 630090, Russia}
\date{\today}

\begin{abstract}
We show that the electron spin polarization generated by an electrical current may have its direction controlled and magnitude amplified by periodic optical excitation. The electrical and optical spin control methods were combined and implemented in a two-dimensional electron gas. By Kerr rotation in an external transverse magnetic field, we demonstrate unexpected long-lived coherent spin oscillations of the current-induced signal in a system with large spin-orbit interaction. Using a single linearly polarized pulse for spin manipulation and detection, we found a strong dependence on the pulse optical power and sample temperature indicating the relevance of the hole spin in the electron spin initialization. The signal was mapped in a Hall bar as function of the position relative to the injection contact. Finally, the presence of an in-plane spin polarization was directly verified by rotating the experimental geometry.
\end{abstract}

\pacs{72.25.Dc, 71.70.Ej, 73.21.Fg, 85.75.-d}

\maketitle
Semiconductor quantum wells are largely explored systems for the future development of spintronic devices using optical or electrical techniques \cite{awschalom_flatte}. For the successful implementation in practical technological platforms, the combination of such methods will be a desirable target for the steps of generation, control with amplification, and detection. Nowadays the reported studies using optical detection, for example Kerr rotation, are mainly divided in two branches depending on the spin polarization trigger used: light or electrical current \cite{examples}.

Among the optical techniques for the generation of spin polarization, those using periodical excitation are very attractive  \cite{glazov_review} because they lead to remarkable phenomena as the resonant spin amplification (RSA) \cite{kikkawa99} and the mode-locking (ML) effect \cite{yugova_rsa,greilich_ML}. At a fixed time delay ($\Delta$t) between a circularly polarized pump and a linearly polarized probe, the amplification of the optically generated out-of-plane component occurs when the spin precession around an external magnetic field (B$_{ext}$) is on resonance with the pulse temporal train structure. In that case, B$_{ext}$ obeys the periodic condition: $\Delta B_{ext} = (h f_{1})/(g \mu_B)$ where $f_{1}$ is the laser repetition rate (see Fig. 1(a) for the RSA geometry). Such synchronization revealed rich spin dynamics for n- and p-doped low-dimensional systems \cite{korn2012} including anisotropic spin relaxation \cite{glazov2008,griesbeck2012}, long-lived spin coherence for electrons and holes in quantum wells \cite{zhukov2012,zhukov2007,korzekwa2013} and quantum dots \cite{spatzek2011,varwig2012,fras2012}, as well as interdependent electron and hole spin dynamics \cite{yugova2009}.

\begin{figure}[h!]
\includegraphics[width=1\columnwidth,keepaspectratio]{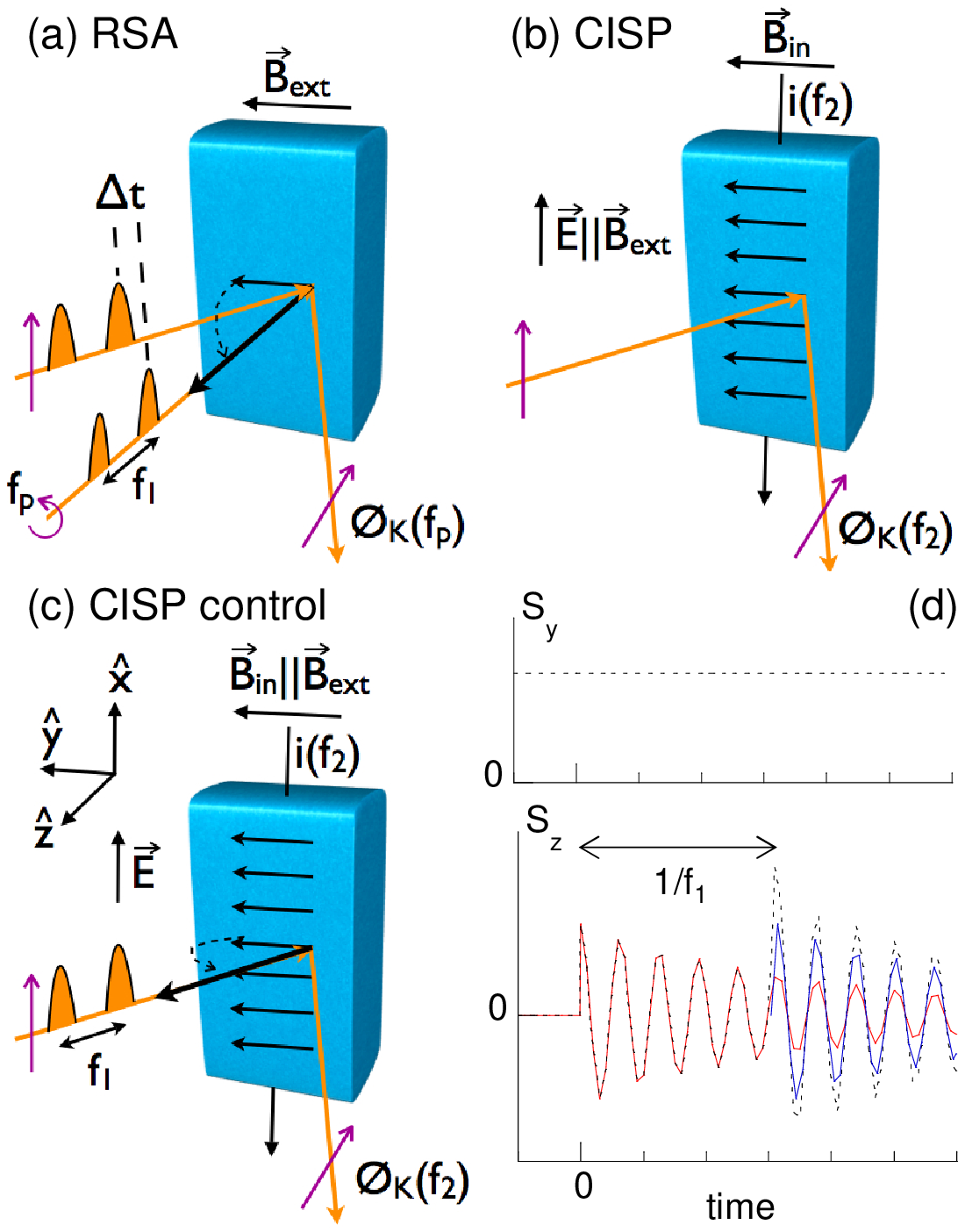}%
\caption{(Color online) (a) Resonant spin amplification. (b) Optical detection of the CISP. (c) Optical control of the CISP. (d) Components of the spin polarization in (c). The Kerr rotation angle ($\phi_{K}$) is detected at the pump modulation frequency ($f_{p}$) or at the applied voltage frequency ($f_2$).}
\end{figure}

On the other side, electrical generation of spin polarization opens a possibility to develop spin current sources \cite{kato2004,stepanov2014} added to the tuning capability of the spin-orbit interaction (SOI) \cite{meier2008,sih2007}. Two components have been observed in two-dimensional electron gases (2DEG): an homogeneous in-plane current-induced spin polarization (CISP) \cite{sih2005}, and an out-of-plane spin polarization accumulated close to the current channel edges resulting from the spin Hall effect (SHE) \cite{sih2005,felix2013}. The CISP in-plane orientation is determined along the internal magnetic field (B$_{in}$) which is tunable by an external electric field (E) and perpendicular to its direction (see Fig. 1(b) for the CISP precession detection around B$_{ext}$).

In a combination of optical and electrical techniques, the application of E on optically generated spin packets showed coherent spin manipulation by the creation of B$_{in}$ \cite{kato2003,kikkawa99_nature}. More recently, the electric field reorientation of also optically generated spins was also demonstrated \cite{kuhlen}. Nevertheless, the opposite situation where a CISP signal is controlled by an optical pulse has not been reported. Here, we address the amplification and reorientation of the CISP by periodic optical excitation in resonance with a variable B$_{ext}$. We explored the manipulation of the spin polarization in a 2DEG by the applied electrical voltage, optical pulse power, and the signal amplitude dependence on the temperature and device geometry. The experimental data displays a RSA pattern with long-lived spin coherence oscillations where the polarization amplitude is enhanced by E, the spin lifetime is independent of the current level but strongly damped by high optical power and temperature.

The experimental geometry is shown in Fig. 1(c). The CISP was optically probed using Kerr rotation (KR). A mode-locked Ti:sapphire laser was used emitting pulses with 100 fs duration at a rate of 76 MHz. The probe laser was linearly polarized and tuned to the absorption edge of the QW sample with adjustable power. The probe beam polarization and amplitude were not modulated and it was detected by coupled photodiodes. The sample was processed in a Hall bar geometry with ohmic contacts for electrical current injection. In order to detect the spin polarization that arises from the electrical pumping, a sine voltage wave with tunable amplitude and fixed frequency of 1.1402 kHz was used for lock-in detection. The sample was immersed in the variable temperature insert of a superconductor magnet in the Voigt geometry with the current channel perpendicular or along B$_{ext}$.

The studied sample is a 45 nm wide GaAs quantum well containing a 2DEG with high electron density n$_{s}$ = 9.2$\times$10$^{11}$ cm$^{-2}$ and mobility of $\mu$ = 1.9$\times$10$^{6}$ cm$^{2}/Vs$ \cite{contact}. The electronic system has a bilayer configuration with symmetric and antisymmetric wave functions for the two lowest subbands and subband separation of $\Delta_{SAS}$ = 1.4 meV. As demonstrated in a previous report \cite{felix2013}, such structure is suitable for the present investigation due to the long spin coherence time comparable with 1/f$_1$.

\begin{figure}[h!]
\includegraphics[width=1\columnwidth,keepaspectratio]{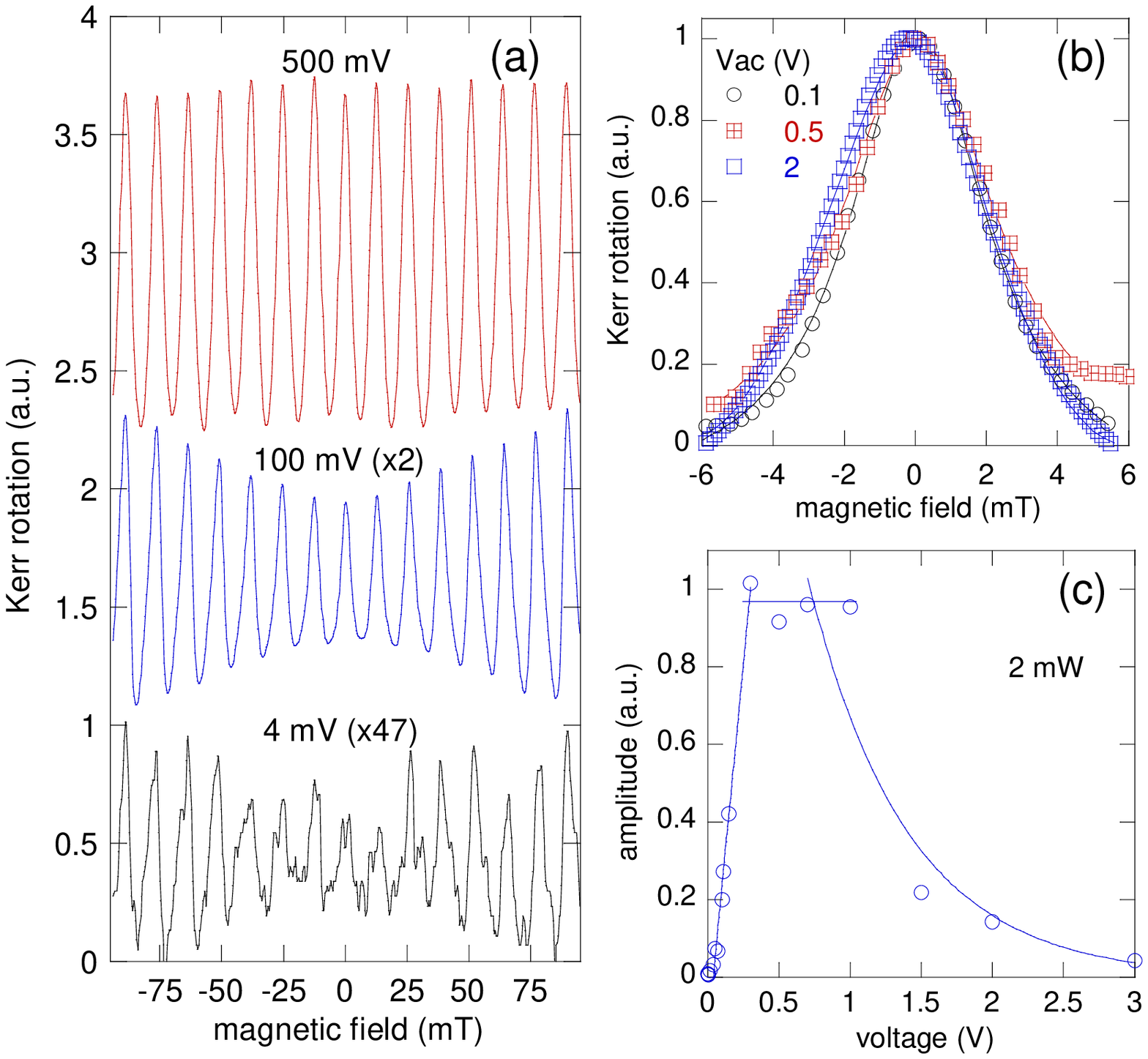}%
\caption{(Color online) Electrical dependence of the CISP manipulation. (a) B$_{ext}$ scans (scaled for clarity) of the KR signal for different applied voltages. (b) Detail of the normalized zero-field peak with a fitted Lorentzian curve for several voltages. (c) KR amplitude dependence on the external voltage.}
\end{figure}

Fig. 2(a) shows the KR signal as function of B$_{ext}$ plotted for several applied voltages in the B$_{ext}$$\perp$E configuration. Unexpectedly, a KR oscillation with field period $\Delta$B$_{ext}$ = 12.5 to 12.8 mT was observed resembling a RSA pattern. By setting  $f_{1}$ = 76 MHz, we get an effective $|g|$ = 0.424 to 0.434 showing an electrically-tunable precession frequency. CISP detection by polar KR was not expected in this B$_{in}$$\parallel$B$_{ext}$ geometry, since the in-plane CISP is aligned with B$_{in}$ (along $\hat{y}$) and thus it can not precess around B$_{ext}$ which leads to zero out-of-plane component. Usually, the B$_{ext}$$\parallel$E configuration is necessary for a direct measurement of the CISP \cite{kato2004} as we shown in Fig. 1(b). We emphasize that there is no additional optical pump beam, and the lock-in detection only uses $f_2$ as reference which acts as the electrical spin pump modulation.

The observation of this remarkable characteristic requires a spin component out of the B$_{ext}$ axis, therefore it implies that pulses in the single linearly polarized beam play a two-fold role to reorient and detect the CISP. Contrary to the generation of spin polarization by circularly polarized light which is related to the angular momentum transfer from the polarized photons, pumping with linearly polarized light should not result in a net spin polarization due to the superposition of left and right circularly polarized components. However, as previously calculated \cite{tarasenko} and observed \cite{schmalbuch}, optical orientation can be produced with linearly polarized light if there is a zero-field spin splitting due to B$_{in}$ induced by the spin-orbit coupling \cite{ganichev} which will lead to asymmetric photoexcitation of carriers with a resulting net spin polarization in the electron gas. The formation of a negatively charged exciton (trion) consisting of an electrically polarized electron spin and optically generated electron-hole pair was examined by the signal dependence on the experimental parameters as discussed below.

The RSA pattern in the CISP signal can be explained as follows. The experiment is initiated by the application of an electrical current which induces spin polarization in the in-plane direction along the B$_{in}$ axis. Fig. 1(d) shows that component ($S_y$) which is constant in time and spatially homogeneous. After electrical generation, a linearly polarized laser pulse reorients the current-induced spin polarization to the out-of-plane direction and generates the $S_z$ component. The first pulse arrives at t=0 and allows the precession of $S_z$ around the total magnetic field (B$_T$) (see also Fig. 1(d)). A pulse of the same laser beam acts as a probe and the lock-in detects the CISP oscillations in KR using the current modulation frequency. Due to the long spin coherence time of the electron system, the pulse temporal train structure leads to the amplification of the $z$ component for fields on resonance. In Fig. 1(d) diagram, the blue and red curves (solid lines) are the $\phi_{K}$ oscillations for successive pulses and the black curves (dashed lines) are the resulting components when B$_{ext}$ gives the constructive condition. This picture agrees with all the presented data in this report.

Another possible explanation could be that signal arises from a purely optical spin generation in a typical RSA with the absorption somehow enhanced in the presence of an electrical current. In that case, the first probe pulse does not reorient the CISP but it creates the spin polarization. This possibility was ruled out by performing an additional experiment using a circularly or linearly polarized pump beam and sweeping the magnetic field at different fixed pump-probe delays. The detection was locked in the optical pump frequency using a photoelastic modulator. With the application of an electric field, the RSA vanishes due to the spin packets drift destroying the amplification condition \cite{SupplementalMaterial}.

Fig. 2(a) shows the enhancement of the RSA pattern amplitude by raising the applied voltage as expected for electrical spin generation \cite{kato2004}. Fig. 2(b) displays a comparison of the zero-field peak for three normalized curves at different voltages. The data points were fitted by a Lorentzian curve $\phi_{K}$=$A/[(\omega_{L}\tau_{s})^{2}+1]$ with half-width $B_{1/2}=\hbar/(g\mu_{B}\tau_{s})$ where A is the KR amplitude, $\omega_{L}=g\mu_{B}B_{ext}/\hbar$ is the Larmor frequency with the electron g-factor $g$, Bohr magneton $\mu_{B}$, Planck's constant $\hbar$, and $\tau_{s}$ is the spin lifetime. In the studied voltage range, the lineshape does not change significantly implying in an approximately constant $\tau_{s}$ = 8 to 10 ns \cite{SupplementalMaterial}. The results for the fitted amplitude are plotted in Fig. 2(c) showing a lever arm of 0.37 between the KR amplitude gain and the applied voltage increment. This linear gain reaches saturation at 0.5 V and reverts to an exponential decay for voltages above 1 V possibly due to current heating. The temperature dependence will be discussed below.

As depicted in Fig. 2 for the voltage dependence, the optical power influence is shown in Fig. 3. Fig. 3(a) displays the lineshape change from broad complex structures to single Lorentzian curves. The unusual RSA profile for low power corresponds to the signal at trion resonance \cite{yugova2009} indicating a strong interdependence of electron and hole dynamics. The increasing RSA amplitude with increasing field for low power implies that faster electron spin precession enhances spin polarization (see also Fig. 2(a)). Such feature was associated with long hole spin relaxation compared to trion recombination (for example, ten times larger) \cite{yugova2009}. Both high current and optical power appear to reduce that ratio which leads to a constant amplitude for increasing B$_{ext}$. The zero-field lineshape with the increasing optical power is plotted in Fig. 3(b) and the fitting results for the amplitude and $\tau_{s}$ are shown in Fig. 3(c). The amplitude increases linearly with the optical power as obtained for the voltage parameter but, in contrast, there is a strong reduction of spin lifetime with high power. In pump-probe experiments \cite{zhukov2007}, the decreasing of $\tau_{s}$ at high pump density was associated with the electrons delocalization caused by their heating due to the interaction with the photogenerated carriers.

\begin{figure}[h]
\includegraphics[width=1\columnwidth,keepaspectratio]{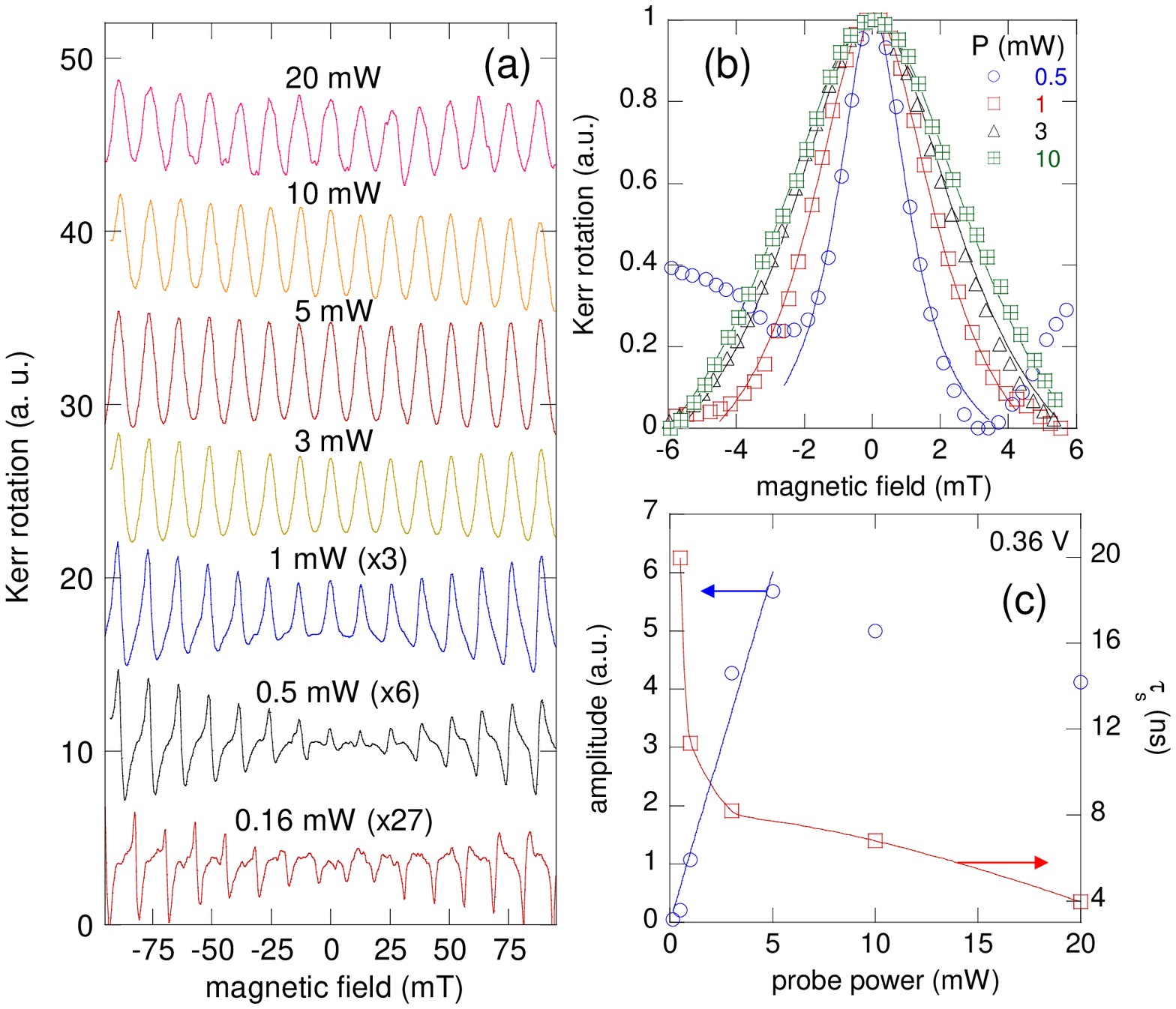}%
\caption{(Color online) Power dependence of the CISP control. (a) B$_{ext}$ scans of the KR signal for different optical power. The curves were scaled for clarity. (b) Detail of the normalized data fit for different power. (c) KR amplitude and spin lifetime dependence for the zero-field peak.}
\end{figure}

\begin{figure}[h!]
\includegraphics[width=1\columnwidth,keepaspectratio]{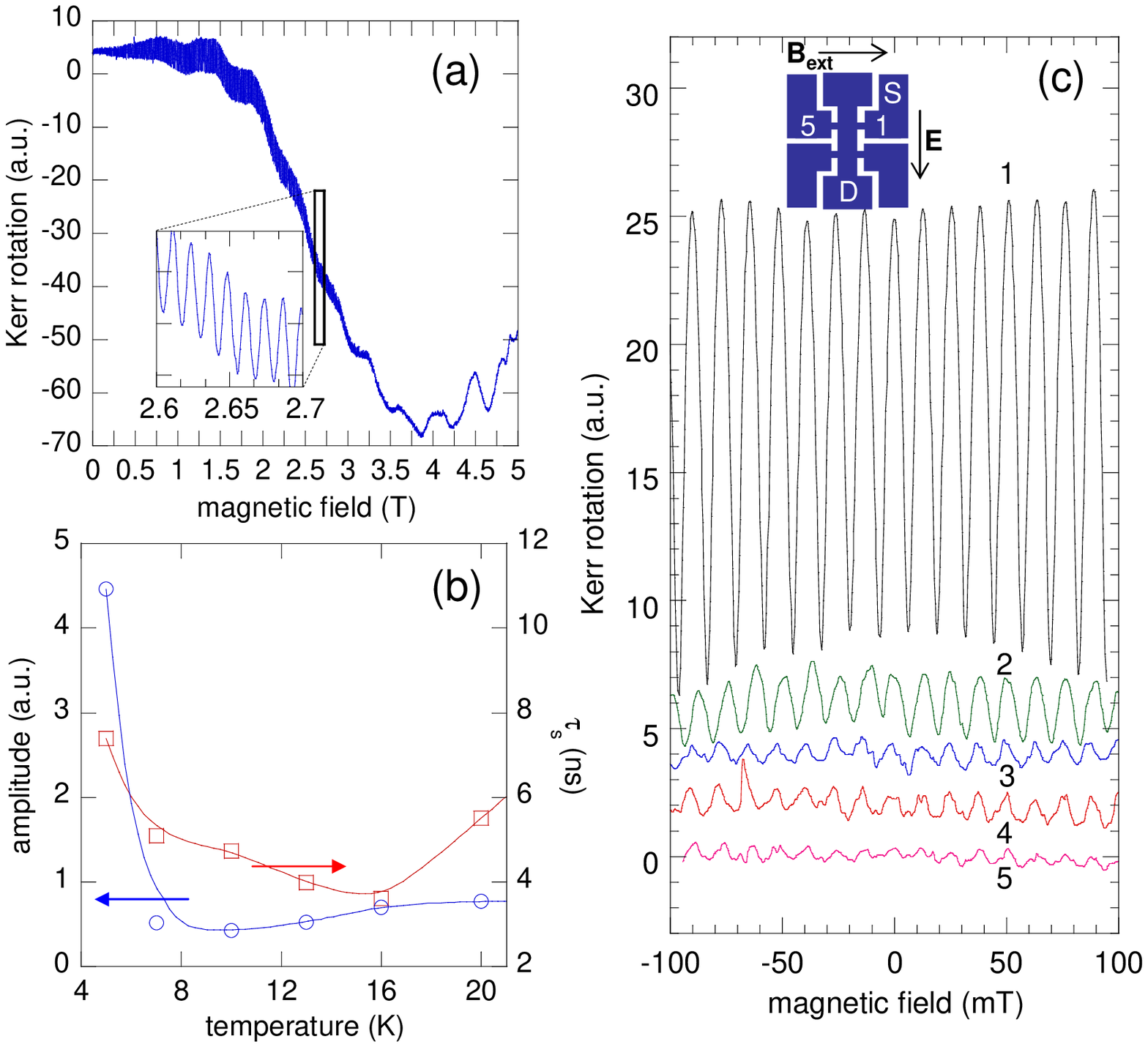}%
\caption{(Color online) CISP optical amplification dependence on (a) magnetic field, (b) temperature, and (c) position relative to the current source and drain contacts.}
\end{figure}

Fig. 4 displays the signal dependence over an extended range of B$_{ext}$, temperature and position in a Hall bar device. In Fig. 4(a), the signal shows a build-up from zero to 0.75 T followed by a slow oscillation with approximate period of 8 T. For the common all-optical RSA, the amplification effect vanishes typically for moderate magnetic field (0.5 to 2 T) \cite{kikkawa99,korn2012,yugova2009} due to the spread in the ensemble electron g-factors which results in an inhomogeneous dephasing. In the present approach, the RSA pattern in Fig. 4(a) holds a constant amplitude up to 3 T and it is strongly suppressed only around 4.25 T. The slow oscillation is associated with the time delay between optical reorientation pump and probe. The measured field period corresponds to $\Delta$t = 20 ps which is similar to the trion recombination lifetime \cite{book_bayer}.

The relevance of current heating, as indicated in Fig. 2(c) and 3(c) decays, was studied by directly raising the temperature up to 20 K.  Although the electron spin lifetime slowly oscillates between 4 and 8 ns, a remarkably strong dependence drops the signal amplitude to about 20\% above 5 K. This observation may be explained by a similar temperature dependence measured for the hole spin relaxation time and related to the spin-orbit coupling for localized holes \cite{yugova2009}. Therefore, the hole spin is a relevant figure in the optical manipulation of the CIPS with dominant contribution in the formation of the trion complex and the resulting electron spin polarization.

An asymmetric current flow was produced in the Hall bar device in order to test the spatial dependence of the CISP amplification. Fig. 4(c) shows the signal measured at five different positions. The current flows from the source contact (S) at the right side of the central channel to the drain at channel bottom (D). The signal is stronger when approaching the injection contact (position 1), approximately constant inside the central channel (2 to 4 positions) and much weaker when close to a region with almost no current flow (position 5). Remarkably, the data shows that the spin polarization induced by the current can be directly mapped by the spatial dependence of the optical amplification level.

Having obtained the spin dynamics key parameters, we demonstrate the experimental method capability as B$_{in}$ sensor. First, we continue in the B$_{ext}$$\perp$E configuration. Fig. 5(a) shows the RSA patterns for several applied voltages where B$_{in}$$\parallel$B$_{ext}$. It is visually clear that the zero-field peak is slightly shifted to negative B$_{ext}$ at 1 V due to the addition of a positive B$_{in}$ \cite{fullrange}. While the central peak for the other two voltages are aligned at B$_{ext}$=0, a larger mismatch is shown by the dashed vertical lines at higher external magnetic field strength, for example at -40 mT, as result of a complementary change in the precession frequency. For larger voltages \cite{felix2013}, the manifestation of the spin Hall effect was observed in a large out-of-plane symmetric polarization peak developed together with the RSA oscillation.

\begin{figure}[h!]
\includegraphics[width=1\columnwidth,keepaspectratio]{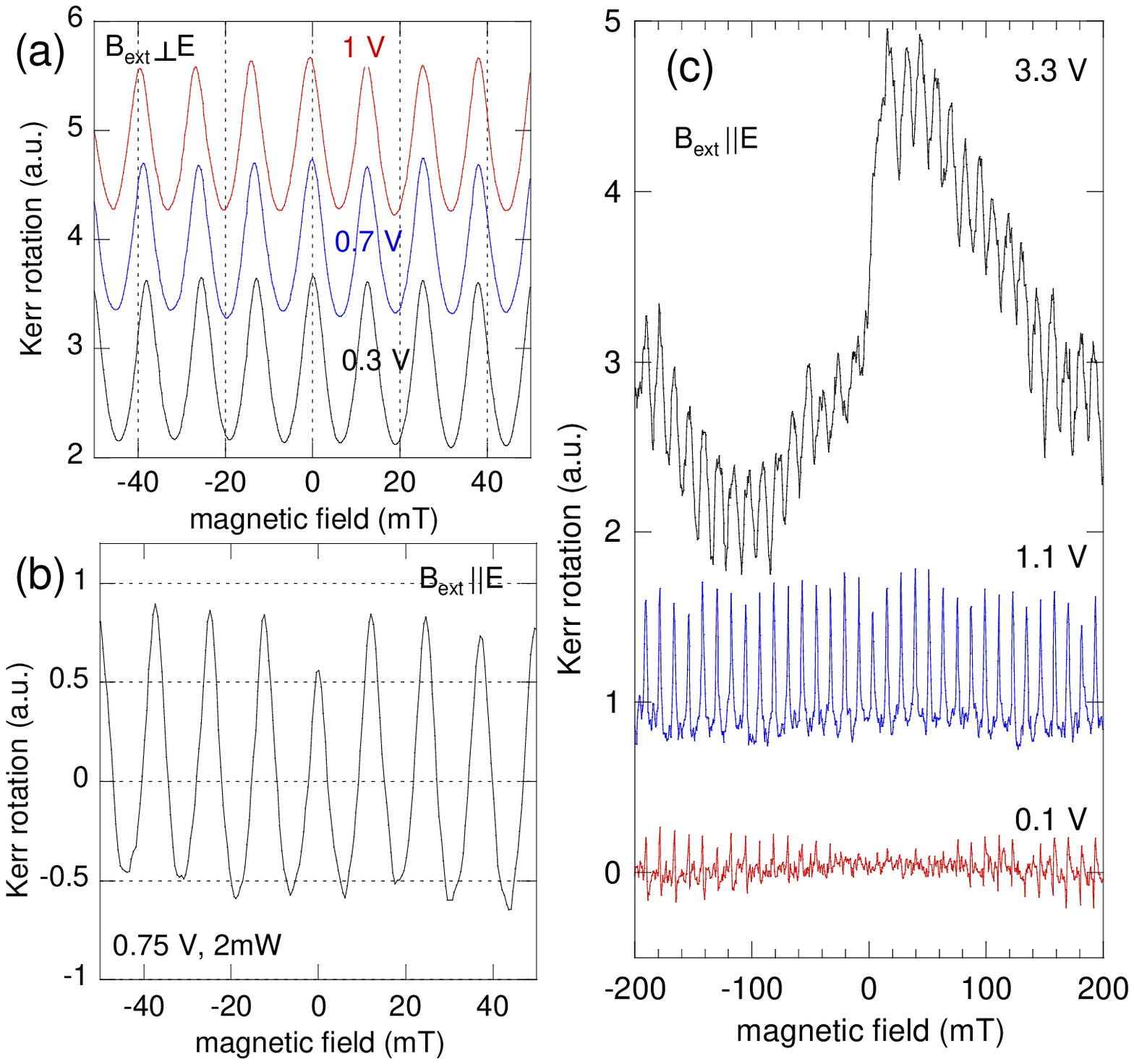}%
\caption{(Color online) Dependence of the RSA pattern on B$_{in}$ for the configuration: (a) B$_{ext}$$\perp$E  and (b) B$_{ext}$$\parallel$E. (c) Direct measurement of the in-plane spin polarization.}
\end{figure}

Finally, we verify the initial assumption of an in-plane CISP in the electron system. To directly measure the spin polarization along $\hat{y}$, we rotate B$_{ext}$ to the $\hat{x}$ axis as shown in Fig. 1(b). The results for the B$_{ext}$$\parallel$E configuration are shown in Fig. 5(b) and Fig. 5(c). The zero-field peak has lower amplitude as expected for an anisotropic relaxation, between the $y$ and $z$ spin components \cite{yugova_rsa,kato2003}, which is electric-field tunable in our experiment. Furthermore, now that B$_{in}$ and B$_{ext}$ are not longer parallel allowing the CISP precession around the magnetic field, the KR signal in Fig. 5(c) displays an antisymmetric Lorenztian curve which is the signature for the CISP in the $\hat{y}$ direction \cite{kato2004}. The RSA pattern is still observed and the resonance sharpness change to broad oscillations as expected for the spin packet drift in a large E.

In summary, we demonstrated the resonant CISP control with a train of optical pulses. We bring together the all-optical RSA and the electrical generation of spin polarization, making possible the study of the polarization components as well as the parameters of the dynamics in a single geometry. Using a single linearly polarized beam for spin orientation and KR detection, the possibility of trion formation consisting of a CISP electron and an optically generated electron-hole pair was examined by the signal dependence on the experimental parameters. The electron and hole spins displayed an interdependent dynamics where the last is crucial for the effective electron polarization as reflected in the oscillations lineshape, B$_{ext}$ and temperature dependence. The RSA pattern revealed a B$_{in}$ and, for B$_{ext}$$\parallel$E, the signal evolves into a antisymmetric polarization peak. The application of the reported effect could provide a path for the implementation of spin current sources induced by charge currents and manipulated by optical ways.

A financial support of this work by grants 2009/15007-5, 2010/09880-5, and 2013/03450-7, S\~{a}o Paulo Research Foundation (FAPESP) is acknowledged. All measurements were done in the LNMS at DFMT-IFUSP.

\end{document}